\documentclass[twocolumn,showpacs,superscriptaddress,amsmath,amssymb,aps,pra]{revtex4-2}
\usepackage{graphicx}
\usepackage{CJKutf8}
\usepackage{dcolumn}
\usepackage{amsmath,bm}
\usepackage{epstopdf}
\usepackage[mathlines]{lineno}
\usepackage{color}
\usepackage[colorlinks=true,linkcolor=blue,citecolor=magenta,urlcolor=cyan]{hyperref}

\usepackage{tikz,xcolor,hyperref}

\definecolor{lime}{HTML}{A6CE39}
\DeclareRobustCommand{\orcidicon}{%
	\begin{tikzpicture}
	\draw[lime, fill=lime] (0,0) 
	circle [radius=0.16] 
	node[white] {{\fontfamily{qag}\selectfont \tiny ID}};
	\draw[white, fill=white] (-0.068,0.105) 
	circle [radius=0.007];
	\end{tikzpicture}
	\hspace{-2mm}
}

\foreach \x in {A, ..., Z}{%
	\expandafter\xdef\csname orcid\x\endcsname{\noexpand\href{https://orcid.org/\csname orcidauthor\x\endcsname}{\noexpand\orcidicon}}
}

\begin{document}
\title{Hole subband dispersions in a cylindrical Ge nanowire: exact results based on the axial Luttinger-Kohn Hamiltonian}
\author{Rui\! Li~(\begin{CJK}{UTF8}{gbsn}李睿\end{CJK})\orcidA{}}
\email{ruili@ysu.edu.cn}
\affiliation{Key Laboratory for Microstructural Material Physics of Hebei Province, School of Science, Yanshan University, Qinhuangdao 066004, China}

\begin{abstract}

Based on the Luttinger-Kohn Hamiltonian in the axial approximation, the transcendental equations determining the hole subband dispersions in a cylindrical Ge nanowire are analytically derived. These equations are more general than that derived using the spherical approximation, and are suitable to study the growth direction dependence of the subband dispersions. The axial approximation almost gives rise to the accurate low-energy subband dispersions for high-symmetry nanowire growth directions [001] and [111]. The perturbation correction from the non-axial term is negligible for these two directions. The lowest two subband dispersions can be regarded as two shifted parabolic curves with an energy gap at $k_{z}=0$ for both growth directions [001] and [111]. At the position of the energy gap, the eigenstates for growth direction [111] are inverted in comparison with the normal eigenstates for growth direction [001]. A nanowire growth direction where the energy gap closes at $k_{z}=0$ is predicted to exist between directions [001] and [111].
\end{abstract}
%\date{October 24, 2023}
\date{\today}
\maketitle

\section{Introduction}

Holes in low dimensional semiconductor nanostructures have attracted considerable interest for decades~\cite{PhysRevB.8.2697,PhysRevB.31.888,ando1985hole,PhysRevB.36.5887,PhysRevB.40.8500,sweeny1988hole,PhysRevB.42.3690,PhysRevB.43.9649,PhysRevB.95.075305}. The early investigations mainly focused on the quasi two-dimensional (2D) hole systems, such as semiconductor quantum wells and heterostructures~\cite{PhysRevB.31.888,ando1985hole,PhysRevB.36.5887,BANGERT1985363}. The effective mass, the subband dispersions, and the spin splitting induced by the bulk and structure inversion asymmetries had been extensively studied~\cite{PhysRevB.31.888,PhysRevB.32.3712,RASHBA1988175,PhysRevB.62.4245,RIDENE201844}. The anisotropic and nonparabolic behaviours of the hole dispersions were revealed~\cite{ando1985hole,PhysRevB.32.5138,PhysRevB.52.11132}, and the Rashba spin splitting was shown to be cubic in momentum~\cite{PhysRevB.62.4245,winkler2003spin}. 

Quasi one-dimensional (1D) hole systems, on the other hand, were relatively less studied. Recently, quasi-1D hole gas was experimentally realized in a Ge/Si core/shell nanowire heterostructure~\cite{Lu10046}, and subsequent experiments revealed a possible strong hole spin-orbit coupling in such system~\cite{PhysRevLett.112.216806,PhysRevResearch.3.013081,Froning:2021aa}. A strong linear in momentum spin-orbit coupling was demonstrated theoretically in the presence of a strong electric field~\cite{PhysRevB.84.195314,PhysRevB.97.235422,PhysRevLett.119.126401} or a strong magnetic field~\cite{RL2022a,RL2023b}. The potential strong spin-orbit coupling has stimulated a series of studies in quasi-1D hole systems, such as the using of hole spins in nanowire quantum dots as quantum information carriers~\cite{PhysRevB.88.241405,Scappucci:2021vk,PhysRevB.105.075308,Wang:2022tm,RL2023a,RL_2024a}, and the searching of Majorana fermions by placing the 1D hole gas in proximity to an s-wave superconductor~\cite{PhysRevB.90.195421,Vries:2018aa}. 

The bulk hole states of semiconductor Ge are described by the $4\times4$ Luttinger-Kohn Hamiltonian~\cite{PhysRev.97.869,PhysRev.102.1030}. It is possible to have an exact solution to the effective mass Hamiltonian for a quasi-2D hole system, e.g., quantum well~\cite{PhysRevB.36.5887}. However, as far as we know, it is impossible for quasi-1D hole system unless approximations are made to the Luttinger-Kohn Hamiltonian. The subband dispersions can be exactly solved in a cylindrical nanowire by using the Luttinger-Kohn Hamiltonian in the spherical approximation~\cite{sweeny1988hole,PhysRevB.42.3690}. However, the spherical approximation fails to account for the growth direction dependence of the subband dispersions. A more reasonable approximation is the axial approximation~\cite{PhysRevB.15.4883,PhysRevB.32.3712,winkler2003spin}. Also, in quasi-2D hole systems, the axial approximation was shown to be most accurate for the high-symmetry growth directions [001] and [111]~\cite{winkler2003spin}.

Here, we extend the exact solvability of the hole subband problem in a cylindrical Ge nanowire from the spherical approximation to the axial approximation. We analytically derive two transcendental equations, the solutions of which give us the hole subband dispersions in the nanowire. The axial approximation is almost accurate for high-symmetry nanowire growth directions [001] and [111]. For example, the first nonvanishing energy correction to the lowest subband dispersion $|F_{z}|=1/2$ comes from its perturbation coupling to the much higher subband dispersion $|F_{z}|=5/2$ (for direction [111]) or $|F_{z}|=7/2$ (for direction [001]). The lowest two subband dispersions can be regarded as two shifted parabolic curves with an energy gap at $k_{z}=0$ for both of the high-symmetry directions. However, the lowest two eigenstates at $k_{z}=0$ for growth direction [111] are inverted in comparison with the normal case for growth direction [001]. We then predict the existence of a gap closing site when the growth direction is rotated from [001] to [111]. Calculations based on the axial approximation confirm this critical growth direction at $\theta\approx38.258^{\circ}$ and $\phi=45^{\circ}$.

\section{Effective mass Hamiltonian of the hole gas}

\begin{figure}
\includegraphics{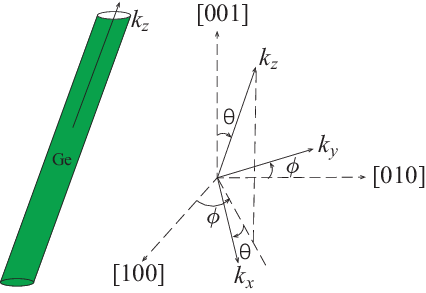}
\caption{\label{fig_coordinate}A cylindrical Ge nanowire grown along the crystallographic direction ($\theta$, $\phi$) is under investigation. The nanowire axis is defined as the $k_{z}$ axis, and the orientations of the coordinate axes $k_{x,y,z}$ relative to the crystal cubic axes are described by two angles $\theta$ and $\phi$. Note that the $k_{y}$ axis is in the [100]-[010] plane.}
\end{figure}

We consider a quasi-1D hole gas in a cylindrical Ge nanowire. The hole gas can move freely in the longitudinal direction while is strongly confined in the transverse direction of the nanowire. The strong transverse confinement leads to the subband quantization of the hole gas~\cite{sweeny1988hole,doi:10.1063/1.4929412,doi:10.1063/1.4972987,PhysRevB.42.3690,ridene2018mid}. For semiconductor Ge, the spin-orbit split-off band is 297 meV far away from the top most valence band~\cite{madelung2004semiconductors}, such that the bulk hole states are well described by the $4\times4$ Luttinger-Kohn Hamiltonian~\cite{PhysRev.102.1030}. Here, in order to study the hole subband dispersions in nanowires grown along various crystallographic directions, we adopt the Luttinger-Kohn Hamiltonian in the axial approximation instead of the frequently used spherical approximation~\cite{PhysRevB.79.155323,PhysRevB.84.195314}. Note that in order to think of the hole energies in the more usual way, we have put an overall minus sign to the hole Hamiltonian. The Luttinger-Kohn Hamiltonian in the axial approximation reads~\cite{winkler2003spin}
\begin{equation}
H^{\rm axial}_{\rm LK}=\frac{\hbar^{2}}{2m}\left(\begin{array}{cccc}
F&S&I&0\\
S^{*}&G&0&I\\
I^{*}&0&G&-S\\
0&I^{*}&-S^{*}&F
\end{array}\right),
\end{equation}
where $m$ is the free electron mass and
\begin{eqnarray}
F&=&(\gamma_{1}+\tilde{\gamma}_{1})(k^{2}_{x}+k^{2}_{y})+(\gamma_{1}-2\tilde{\gamma}_{1})k^{2}_{z},\nonumber\\
G&=&(\gamma_{1}-\tilde{\gamma}_{1})(k^{2}_{x}+k^{2}_{y})+(\gamma_{1}+2\tilde{\gamma}_{1})k^{2}_{z},\nonumber\\
S&=&-2\sqrt{3}\tilde{\gamma}_{2}k_{z}k_{-},\nonumber\\
I&=&-\sqrt{3}\tilde{\gamma}_{3}k^{2}_{-},
\end{eqnarray}
with $k_{\pm}=k_{x}\pm\,ik_{y}$ and 
\begin{eqnarray}
\tilde{\gamma}_{1}&=&(1-\zeta)\gamma_{2}+\zeta\gamma_{3},\nonumber\\
\tilde{\gamma}_{2}&=&\frac{2}{3}\zeta\gamma_{2}+(1-\frac{2}{3}\zeta)\gamma_{3},\nonumber\\
\tilde{\gamma}_{3}&=&\frac{1}{6}(3-\zeta)\gamma_{2}+\frac{1}{6}(3+\zeta)\gamma_{3}.\label{eq_gamma}
\end{eqnarray}
Here $\zeta=\sin^{2}\theta[3-(3/8)(7+\cos4\phi)\sin^{2}\theta]$, where $\theta$ is the azimuthal angle and $\phi$ is the polar angle of the nanowire axis with respect to the [001] direction (see Fig.~\ref{fig_coordinate})~\cite{winkler2003spin}, and $\gamma_{1}=13.35$, $\gamma_{2}=4.25$, and $\gamma_{3}=5.69$ are the Luttinger parameters for semiconductor Ge~\cite{PhysRevB.4.3460}. 

We use an infinite cylindrical well to model the transverse confining potential of the hole gas, such that the effective mass Hamiltonian of the hole gas reads (the real hole Hamiltonian should be interpreted as $-H$)
\begin{equation}
H=H^{\rm axial}_{\rm LK}+V(r),\label{eq_Hamiltonain1}
\end{equation}
where
\begin{equation}
V(r)=\left\{\begin{array}{cc}0,~&~r<R,\\
\infty,~&~r>R,\end{array}\right.
\end{equation}
with $r=\sqrt{x^{2}+y^{2}}$ and $R$ being the radius of the Ge nanowire. We note that in the effective mass model (\ref{eq_Hamiltonain1}), the replacements $k_{x,y}=-i\partial_{x,y}$ should be made. Interestingly, Hamiltonian $H$ commutates with the total angular momentum $F_{z}=-i\partial_{\varphi}+J_{z}$, where $\varphi=\arctan(x/y)$ and $J_{z}$ is the $z$ component of the standard spin $3/2$ matrices~\cite{winkler2003spin}. Hence, Hamiltonian $H$ and operator $F_{z}$ have common eigenstates. This is the key to the exact solution of the effective mass model (\ref{eq_Hamiltonain1}).

Although the Luttinger-Kohn Hamiltonian in the axial approximation is adopted in this paper, in the following we will show that the axial approximation is indeed an excellent approximation for the high-symmetry nanowire growth directions [001] ($\theta=0^{\circ}$ and $\phi=0^{\circ}$) and [111] ($\theta=\arccos(1/\sqrt{3})$ and $\phi=45^{\circ}$). This assertation is also well known in the quasi-2D hole subband calculations, where the growth direction refers to a quantum well~\cite{winkler2003spin}.

\section{The transcendental equations}

For a given value of $k_{z}$, we want to solve both the eigenvalues and the corresponding eigenfunctions of Hamiltonian (\ref{eq_Hamiltonain1}). Due to the conservation of the total angular momentum $F_{z}$~\cite{winkler2003spin}, the eigenfunctions can be classified by $F_{z}=m+1/2$, with $m$ being an integer. Inside the well $r<R$, the confining potential is zero, Hamiltonian $H$ is reduced to the bulk Hamiltonian $H^{\rm axial}_{\rm LK}$. Therefore, our first step is to obtain both the bulk spectrum and the corresponding bulk wavefunctions, i.e., the common eigenfunctions of $H^{\rm axial}_{\rm LK}$ and $F_{z}$. Then, we expand the eigenfunction in terms of the bulk wavefunctions by introducing a series of coefficients. At last, letting the eigenfunction satisfy the desired hard-wall boundary condition, we are able to obtain an equation array for the coefficients. Solving this equation array leads to the exact solution of the effective mass model (\ref{eq_Hamiltonain1}). 

\begin{figure}
\includegraphics{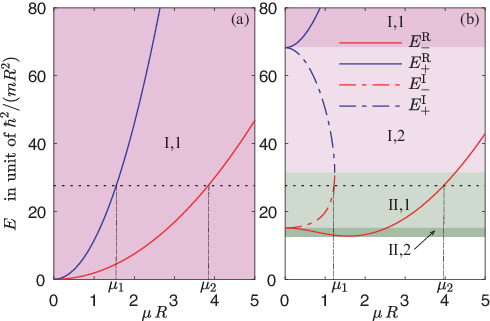}
\caption{\label{fig_bulkspectrum}The bulk hole dispersions for growth direction [001] with the longitudinal wave vector fixed at $k_{z}R=0$ (a) and $k_{z}R=2.5$ (b). We can specify the whole energy region as region I when $k_{z}R=0$ or divide it into two regions I and II when $k_{z}R\neq0$. A constant energy line in the energy region I intersects the plus branch dispersion (either $E^{\rm R}_{+}$ or $E^{\rm I}_{+}$) at $\mu_{1}$, and intersects the minus branch dispersion $E^{\rm R}_{-}$ at $\mu_{2}$. While a constant energy line in the energy region II intersects the minus branch dispersion (either $E^{\rm R}_{-}$ or $E^{\rm I}_{-}$) both at $\mu_{1}$ and $\mu_{2}$. }
\end{figure}

The bulk spectrum and the corresponding bulk wavefunctions can be obtained by solving $H^{\rm axial}_{\rm LK}\Psi_{b}(r,\varphi,z)=E\Psi_{b}(r,\varphi,z)$ in the cylindrical coordinate system where $x=r\cos\varphi$, $y=r\sin\varphi$, and $z=z$. The detailed derivations are given in appendix~\ref{app_a}, we have two branches of bulk spectrum in terms of the Bessel functions $J_{m}(\mu\,r)$~\cite{zhuxi_wang}
\begin{equation}
E^{\rm R}_{\pm}=\frac{\hbar^{2}}{2m}[\gamma_{1}(\mu^{2}+k^{2}_{z})\pm\,X_{\mu}],\label{eq_bulkspectrum}
\end{equation}
where
\begin{equation}
X_{\mu}=\sqrt{\tilde{\gamma}^{2}_{1}(\mu^{2}-2k^{2}_{z})^{2}+3\mu^{2}(4\tilde{\gamma}^{2}_{2}k^{2}_{z}+\tilde{\gamma}^{2}_{3}\mu^{2})},\label{eq_Xmu}
\end{equation}
with $\mu^{2}=k^{2}_{x}+k^{2}_{y}$. For each branch of bulk spectrum, there exist two bulk wavefunctions whose explicit expressions are also given in appendix~\ref{app_a}. In order to solve the subband energies in the whole energy region $[E^{\rm R}_{\rm min},\infty]$, where $E^{\rm R}_{\rm min}$ is the energy minimum of the minus dispersion $E^{\rm R}_{-}$, we also need to utilize the modified Bessel functions $J_{m}(i\mu\,r)$, i.e., the Bessel function with imaginary argument $i\mu\,r$.  A simple replacement of $\mu$ with $i\mu$ in Eq.~(\ref{eq_bulkspectrum}) gives us the bulk hole spectrum in terms of the modified Bessel functions~\cite{zhuxi_wang}
\begin{equation}
E^{\rm I}_{\pm}=\left.E^{\rm R}_{\pm}\right|_{\mu=i\mu}.
\end{equation}
The bulk hole dispersions for growth direction [001] are shown in Fig.~\ref{fig_bulkspectrum}. We can specify the whole energy region as region I when $k_{z}R=0$ or divide it into two regions I and II when $k_{z}R\neq0$ (see Fig.~\ref{fig_bulkspectrum}).

The eigenfunction of Hamiltonian (\ref{eq_Hamiltonain1}) can be expanded in terms of the four bulk wave functions with respect to the energy region. In energy region I, we should choose two from the `+' branch given by Eqs.~(\ref{eq_bulkstateplus1}) and (\ref{eq_bulkstateplus2}), and two from the `-' branch given by Eqs.~(\ref{eq_bulkstateminus1}) and (\ref{eq_bulkstateminus2}). While in energy region II, all the four bulk wave functions should be chosen from the `-' branch given by Eqs.~(\ref{eq_bulkstateminus1}) and (\ref{eq_bulkstateminus2}). Therefore, the eigenfunction $\Psi^{\rm I/II}(r,\varphi,z)$ in energy region I/II can be written as 
\begin{eqnarray}
\Psi^{\rm I/II}(r,\varphi,z)&\equiv&\left(\begin{array}{c}\Psi_{1}^{\rm I/II}(r)e^{i(m-1)\varphi}\\\Psi_{2}^{\rm I/II}(r)e^{im\varphi}\\\Psi_{3}(r)e^{i(m+1)\varphi}\\\Psi_{4}(r)e^{i(m+2)\varphi}\end{array}\right)e^{ik_{z}z},\label{Eq_eigenfunc}
\end{eqnarray}
where
\begin{widetext}
\begin{eqnarray}
\Psi_{1}^{\rm I/II}(r)&=&\left(c_{1}\frac{2i\tilde{\gamma}_{2}k_{z}}{\tilde{\gamma}_{3}\mu_{1}}+c_{2}\frac{\tilde{\gamma}_{1}(\mu^{2}_{1}-2k^{2}_{z})\pm\,X_{\mu_{1}}}{\sqrt{3}\tilde{\gamma}_{3}\mu^{2}_{1}}\right)J_{m-1}(\mu_{1}\,r)+\left(c_{3}\frac{2i\tilde{\gamma}_{2}k_{z}}{\tilde{\gamma}_{3}\mu_{2}}+c_{4}\frac{\tilde{\gamma}_{1}(\mu^{2}_{2}-2k^{2}_{z})-X_{\mu_{2}}}{\sqrt{3}\tilde{\gamma}_{3}\mu^{2}_{2}}\right)J_{m-1}(\mu_{2}\,r),\nonumber\\
\Psi_{2}^{\rm I/II}(r)&=&\left(c_{1}\frac{-\tilde{\gamma}_{1}(\mu^{2}_{1}-2k^{2}_{z})\pm\,X_{\mu_{1}}}{\sqrt{3}\tilde{\gamma}_{3}\mu^{2}_{1}}-c_{2}\frac{2i\tilde{\gamma}_{2}k_{z}}{\tilde{\gamma}_{3}\mu_{1}}\right)J_{m}(\mu_{1}\,r)+\left(c_{3}\frac{-\tilde{\gamma}_{1}(\mu^{2}_{2}-2k^{2}_{z})-X_{\mu_{2}}}{\sqrt{3}\tilde{\gamma}_{3}\mu^{2}_{2}}-c_{4}\frac{2i\tilde{\gamma}_{2}k_{z}}{\tilde{\gamma}_{3}\mu_{2}}\right)J_{m}(\mu_{2}\,r),\nonumber\\
\Psi_{3}(r)&=&c_{2}J_{m+1}(\mu_{1}\,r)+c_{4}J_{m+1}(\mu_{2}\,r),\nonumber\\
\Psi_{4}(r)&=&c_{1}J_{m+2}(\mu_{1}\,r)+c_{3}J_{m+2}(\mu_{2}\,r),\label{eq_wavefunction}
\end{eqnarray}
\end{widetext}
with $J_{m}(\mu\,r)$ being the $m$-order Bessel function~\cite{zhuxi_wang}, and
\begin{equation}
\mu_{1,2}=\left(\frac{-b\mp\sqrt{b^{2}-4ac}}{2a}\right)^{1/2},\label{eq_mu}
\end{equation}
with
\begin{eqnarray}
a&=&\gamma^{2}_{1}-\tilde{\gamma}^{2}_{1}-3\tilde{\gamma}^{2}_{3},\nonumber\\
b&=&2\big[\gamma^{2}_{1}k^{2}_{z}+2(\tilde{\gamma}^{2}_{1}-3\tilde{\gamma}^{2}_{2})k^{2}_{z}-2\gamma_{1}mE/\hbar^{2}\big],\nonumber\\
c&=&\gamma^{2}_{1}k^{4}_{z}-4\tilde{\gamma}^{2}_{1}k^{4}_{z}-4\gamma_{1}k^{2}_{z}mE/\hbar^{2}+4m^{2}E^{2}/\hbar^{4}.
\end{eqnarray}
Here $\mu_{1,2}$ are obtained by inversely solving Eq.~(\ref{eq_bulkspectrum}). Note that $\mu_{2}$ is always real in the whole energy region (see Fig.~\ref{fig_bulkspectrum}), while $\mu_{1}$ can be real in the energy subregions I,1 and II,2 or imaginary in the energy subregions I,2 and II,1 [see Fig.~\ref{fig_bulkspectrum}(b)]. Because $\mu_{1}$ is naturally real or imaginary in different energy subregions, such that the modified Bessel functions are also naturally incorporated in Eq.~(\ref{eq_wavefunction}) .

Imposing the hard-wall boundary condition to the eigenfunction $\Psi^{\rm I/II}(R,\varphi,z)=0$, we obtain four equations for the expansion coefficients $c_{1,2,3,4}$ (for details see appendix~\ref{app_b}). The determinant of the coefficient matrix must be zero, such that we have the following transcendental equations 
\begin{widetext}
\begin{equation}
\left|\begin{array}{cccc}
\frac{2i\tilde{\gamma}_{2}k_{z}}{\tilde{\gamma}_{3}\mu_{1}}J_{m-1}(\mu_{1}R)&\frac{\tilde{\gamma}_{1}(\mu^{2}_{1}-2k^{2}_{z})\pm\,X_{\mu_{1}}}{\sqrt{3}\tilde{\gamma}_{3}\mu^{2}_{1}}J_{m-1}(\mu_{1}R)&\frac{2i\tilde{\gamma}_{2}k_{z}}{\tilde{\gamma}_{3}\mu_{2}}J_{m-1}(\mu_{2}R)&\frac{\tilde{\gamma}_{1}(\mu^{2}_{2}-2k^{2}_{z})-X_{\mu_{2}}}{\sqrt{3}\tilde{\gamma}_{3}\mu^{2}_{2}}J_{m-1}(\mu_{2}R)\\
\frac{-\tilde{\gamma}_{1}(\mu^{2}_{1}-2k^{2}_{z})\pm\,X_{\mu_{1}}}{\sqrt{3}\tilde{\gamma}_{3}\mu^{2}_{1}}J_{m}(\mu_{1}R)&-\frac{2i\tilde{\gamma}_{2}k_{z}}{\tilde{\gamma}_{3}\mu_{1}}J_{m}(\mu_{1}R)&\frac{-\tilde{\gamma}_{1}(\mu^{2}_{2}-2k^{2}_{z})-X_{\mu_{2}}}{\sqrt{3}\tilde{\gamma}_{3}\mu^{2}_{2}}J_{m}(\mu_{2}R)&-\frac{2i\tilde{\gamma}_{2}k_{z}}{\tilde{\gamma}_{3}\mu_{2}}J_{m}(\mu_{2}R)\\
0&J_{m+1}(\mu_{1}R)&0&J_{m+1}(\mu_{2}R)\\
J_{m+2}(\mu_{1}R)&0&J_{m+2}(\mu_{2}R)&0
\end{array}\right|=0.\label{eq_transc}
\end{equation}
\end{widetext}
Note that there is a `$\pm$'$\rightarrow$I/II correspondence in the above equations, i.e., the plus sign `$+$' and the minus sign `-' respectively give rise to the transcendental equation in the energy regions I and II. These two equations are implicit equations of the energy eigenvalue $E$. By fixing the total angular momentum $F_{z}=m+1/2$, i.e., fixing $m$, for a given value of $k_{z}$, we can solve a series of energies satisfying Eq.~(\ref{eq_transc}). When the energies are plotted as a function of $k_{z}$, we have the subband dispersions of the hole gas. Once an eigenvalue $E$ is solved from Eq.~(\ref{eq_transc}), we can solve the coefficients $c_{1,2,3,4}$ from the matrix equation (see appendix~\ref{app_b}) in combination with the normalization condition of the eigenfunction~\cite{RL2021}. %The two components $\Psi_{1,3}(r)$ of the eigenfunction $\Psi(r,\varphi,z)$ can be chosen as imaginary, while the other two components $\Psi_{2,4}(r)$ can be chosen as real.

We note that from symmetry analysis~\cite{RL2021}, the hole energies are two-fold degenerate, i.e., spin degenerate. If Eq.~(\ref{Eq_eigenfunc}) is an eigenfunction with eigenvalue $E(k_{z})$, then the same is the following eigenfunction 
\begin{eqnarray}
\left(\begin{array}{c}\Psi^{*}_{4}(r)e^{-i(m+2)\varphi}\\\Psi^{*}_{3}(r)e^{-i(m+1)\varphi}\\\Psi^{*}_{2}(r)e^{-im\varphi}\\\Psi^{*}_{1}(r)e^{-i(m-1)\varphi}\end{array}\right)e^{ik_{z}z}.\label{Eq_eigenfunc2}
\end{eqnarray}
Eigenfunction (\ref{Eq_eigenfunc2}) corresponds to total angular momentum $F_{z}=-(m+1/2)$. Hence, combining Eqs.~(\ref{Eq_eigenfunc}) and (\ref{Eq_eigenfunc2}), we can simply write the hole spin degeneracy in the subbands as $|F_{z}|=m+1/2$, with $m=0,1,2...$.

\subsection{The case $k_{z}=0$}

We discuss here the subband energies and the corresponding eigenfunctions at the special wave vector site $k_{z}=0$~\cite{PhysRevB.79.155323}. When $k_{z}=0$, the energy region II disappears [see Fig.~\ref{fig_bulkspectrum}(a)], and the plus sign `+' transcendental equation (\ref{eq_transc}) can be reduced to two independent equations, one reads
\begin{eqnarray}
&&\frac{\sqrt{\tilde{\gamma}^{2}_{1}+3\tilde{\gamma}^{2}_{3}}+\tilde{\gamma}_{1}}{\sqrt{\tilde{\gamma}^{2}_{1}+3\tilde{\gamma}^{2}_{3}}-\tilde{\gamma}_{1}}J_{m-1}(\mu_{1}R)J_{m+1}(\mu_{2}R)\nonumber\\
&&~~~~~~+J_{m-1}(\mu_{2}R)J_{m+1}(\mu_{1}R)=0,\label{eq_transc1}
\end{eqnarray}
and the other reads
\begin{eqnarray}
&&\frac{\sqrt{\tilde{\gamma}^{2}_{1}+3\tilde{\gamma}^{2}_{3}}+\tilde{\gamma}_{1}}{\sqrt{\tilde{\gamma}^{2}_{1}+3\tilde{\gamma}^{2}_{3}}-\tilde{\gamma}_{1}}J_{m}(\mu_{2}R)J_{m+2}(\mu_{1}R)\nonumber\\
&&~~~~~~+J_{m}(\mu_{1}R)J_{m+2}(\mu_{2}R)=0.\label{eq_transc2}
\end{eqnarray}
If the energy eigenvalue $E$ is solved from Eq.~(\ref{eq_transc1}), we are easy to find a simple solution for the expansion coefficients
\begin{equation}
c_{2}=iJ_{m+1}(\mu_{2}R),~c_{4}=-iJ_{m+1}(\mu_{1}R),~c_{1}=c_{3}=0.\label{eq_coefficient1}
\end{equation}
This result indicates that two components of the eigenfunction [see Eq.~(\ref{eq_wavefunction})] are zero, i.e., $\Psi_{2}(r)=\Psi_{4}(r)=0$. If the energy eigenvalue $E$ is solved from Eq.~(\ref{eq_transc2}), the expansion coefficients of the eigenfunction can be solved as
\begin{equation}
c_{1}=J_{m+2}(\mu_{2}R),~c_{3}=-J_{m+2}(\mu_{1}R),~c_{2}=c_{4}=0.\label{eq_coefficient2}
\end{equation}
This result indicates that the following two components $\Psi_{1}(r)$ and $\Psi_{3}(r)$ of the eigenfunction are zero instead. 

Hence, at the site $k_{z}=0$, the eigenfunctions always have two vanishing components, i.e., either $\Psi_{1,3}(r)$ or $\Psi_{2,4}(r)$ are zero~\cite{PhysRevB.79.155323,PhysRevB.84.195314,RL2021}. We also note that by choosing the simple solutions as Eqs.~(\ref{eq_coefficient1}) and (\ref{eq_coefficient2}), the eigenfunctions are not normalized.

\subsection{The case $\tilde{\gamma}_{1}=\tilde{\gamma}_{2}=\tilde{\gamma}_{3}$}

Here, we temporarily ignore the expressions of $\tilde{\gamma}_{1,2,3}$ given in Eq.~(\ref{eq_gamma}), and consider the ideal case where $\tilde{\gamma}_{1}=\tilde{\gamma}_{2}=\tilde{\gamma}_{3}=\gamma_{s}$. In this case, the Luttinger-Kohn Hamiltonian in the axial approximation $H^{\rm axial}_{\rm LK}$ is reduced to that in the spherical approximation. The formulas based on the spherical approximation in calculating the hole subband dispersions are well established in the literature~\cite{sweeny1988hole,PhysRevB.42.3690}. 

Now, the parameter $X_{\mu}$ defined in Eq.~(\ref{eq_Xmu}) can be simplified to
\begin{equation}
X_{\mu}=2\gamma_{s}(\mu^{2}+k^{2}_{z}),
\end{equation}
and the $\mu_{1,2}$ given in Eq.~(\ref{eq_mu}) can be simplified to
\begin{equation}
\mu_{1,2}=\sqrt{\frac{2mE}{(\gamma_{1}\pm2\gamma_{s})\hbar^{2}}-k^{2}_{z}}.
\end{equation}
Substituting these results into the the plus sign `+' transcendental equation Eq.~(\ref{eq_transc}) , we recover the well-known transcendental equation given by Sercel and Vahala~\cite{PhysRevB.42.3690}. Note that in the spherical approximation, the energy region II disappears. Also, the two independent transcendental equations (\ref{eq_transc1}) and (\ref{eq_transc2}) at $k_{z}=0$ can be reduced to that given previously~\cite{PhysRevB.79.155323,PhysRevB.84.195314}.

\section{Subband dispersions and subband wavefunctions}

\subsection{The results for growth direction [001]}

\begin{figure}
\includegraphics{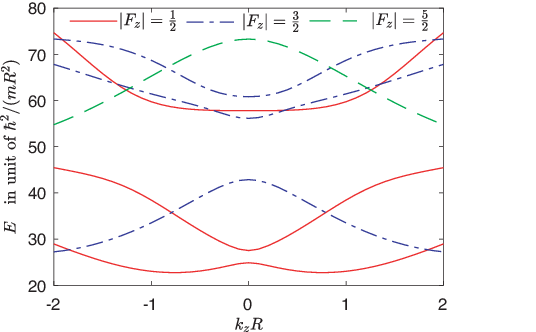}
\caption{\label{fig_subbands001}Hole subband dispersions for growth direction [001]. For a moderate nanowire radius $R=10$ nm, the energy unit is $\hbar^{2}/(mR^{2})\approx0.763$ meV. The energy gap at the anticrossing site $k_{z}R=0$ of the lowest two subband dispersions is about $2.63\hbar^{2}/(mR^{2})$. The band minima are located at $k_{z}R\approx\pm0.76$. Each dispersion line is two-fold degenerate, i.e., spin degenerate.}
\end{figure}

When the growth direction of the Ge nanowire is along the [001] crystal axis, i.e., $k_{x}\parallel[100]$, $k_{y}\parallel[010]$, and $k_{z}\parallel[001]$, the parameters $\tilde{\gamma}_{1,2,3}$ in the Luttinger-Kohn Hamiltonian in the axial approximation read $\tilde{\gamma}_{1}=\gamma_{2}$, $\tilde{\gamma}_{2}=\gamma_{3}$, and $\tilde{\gamma}_{3}=(\gamma_{2}+\gamma_{3})/2$~\cite{PhysRevB.52.11132}. Substituting these parameter values into Eq.~(\ref{eq_transc}), we can solve the hole subband dispersions for this growth direction. 

The hole subband dispersions for growth direction [001] are shown in Fig.~\ref{fig_subbands001}. Each line in the figure is two-fold degenerate, i.e., spin degenerate. The lowest two subband dispersions, i.e., the lowest two solid lines of $|F_{z}|=1/2$, can be regarded as two shifted parabolic curves with an anticrossing at $k_{z}R=0$~\cite{RL2022a,RL2023b,RL2023c}. The energy gap at the anticrossing site $k_{z}R=0$ is about $2.63\hbar^{2}/(mR^{2})$. The band minima are located at $k_{z}R\approx\pm0.76$. We note that the lowest three subband dispersions shown in Fig.~\ref{fig_subbands001} agree well with that obtained previously using the perturbation method~\cite{RL2023c}. We also give the lowest subband dispersions both of $|F_{z}|=3/2$ and $|F_{z}|=5/2$ for wave vectors up to $|k_{z}R|=5$ in appendix~\ref{app_c}, where the minima of these two subband dispersions are covered now.

\begin{figure}
\includegraphics{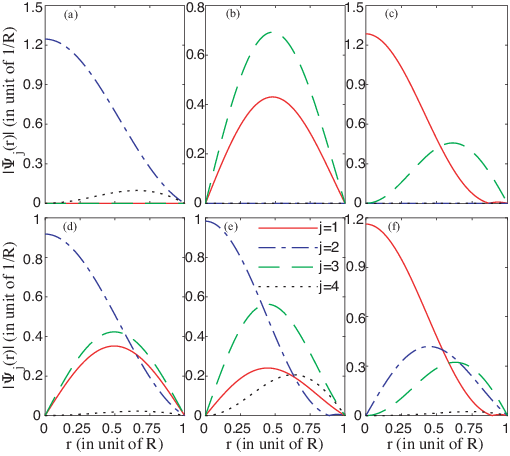}
\caption{\label{fig_subbandwfs001}Hole subband wavefunctions for growth direction [001]. The four components $\Psi_{1,2,3,4}(r)$ of the wavefunction Eq.~(\ref{Eq_eigenfunc}) plotted as a function of the coordinate $r$.  The ground (a), the first excited (b), and the second excited (c) states at $k_{z}R=0$. The ground (d), the first excited (e), and the second excited (f) states at $k_{z}R=0.8$. Here, we only consider eigenfunctions with positive total angular momentum $F_{z}>0$.}
\end{figure}

The functional forms of the subband wavefunctions are shown in Fig.~\ref{fig_subbandwfs001}. At the center of the wave vector space $k_{z}R=0$, the wavefunctions always have two non-zero components. Figures~\ref{fig_subbandwfs001}(a), (b) and (c) show the results at $k_{z}R=0$ of the ground, the first excited, and the second excited states, respectively. Here the ground state has components $\Psi_{2,4}(r)\neq0$ and the first excited state has components $\Psi_{1,3}(r)\neq0$.  At the site $k_{z}R=0.8$, all the four components of the eigenfunction are nonzero, see the results of the ground, the first excited, and the second excited states in Figs.~\ref{fig_subbandwfs001}(d), (e) and (f), respectively. Here we only show the eigenfunctions with positive total angular momentum $F_{z}>0$.

\subsection{The results for growth direction [111]}

\begin{figure}
\includegraphics{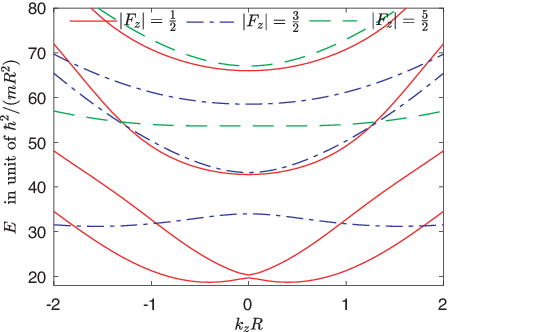}
\caption{\label{fig_subbands111}Hole subband dispersions for growth direction [111]. The energy gap at the anticrossing site $k_{z}R=0$ of the lowest two subband dispersions is about $0.61\hbar^{2}/(mR^{2})$. The band minima are located at $k_{z}R\approx\pm0.4$. Each dispersion line is two-fold degenerate, i.e., spin degenerate. }
\end{figure}

When the growth direction of the Ge nanowire is along the [111] crystal axis, i.e., $k_{x}\parallel[11\bar{2}]$, $k_{y}\parallel[\bar{1}10]$, and $k_{z}\parallel[111]$, the parameters $\tilde{\gamma}_{1,2,3}$ in the Luttinger-Kohn Hamiltonian in the axial approximation read $\tilde{\gamma}_{1}=\gamma_{3}$, $\tilde{\gamma}_{2}=(2\gamma_{2}+\gamma_{3})/3$, and $\tilde{\gamma}_{3}=(\gamma_{2}+2\gamma_{3})/3$~\cite{PhysRevB.85.235308,PhysRevB.108.165301}. Substituting these parameter values into the transcendental equation (\ref{eq_transc}), we can solve the hole subband dispersions for this growth direction. 

The subband dispersions for growth direction [111] are shown in Fig.~\ref{fig_subbands111}. Each dispersion line in the figure is two-fold degenerate, e.g. spin degenerate. The lowest two subband dispersions, i.e., the lowest two solid lines of $|F_{z}|=1/2$, can still be regarded as two shifted parabolic curves with an anticrossing at $k_{z}R=0$~\cite{RL2022a,RL2023b,RL2023c}. The energy gap at the anticrossing site is about $0.61\hbar^{2}/(mR^{2})$, much smaller than that for growth direction [001]. The band minima are located at $k_{z}R\approx\pm0.40$, much closer to the center of $k_{z}$ space than that for growth direction [001]. 

\begin{figure}
\includegraphics{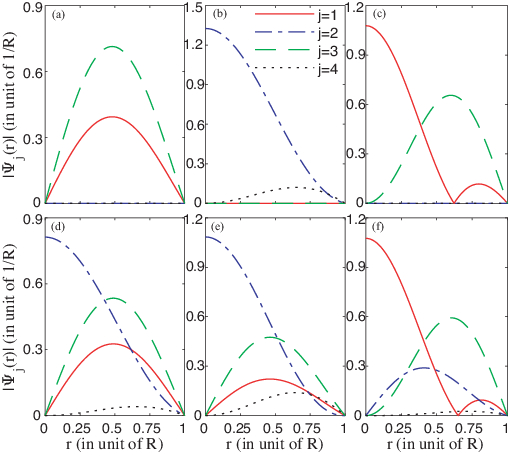}
\caption{\label{fig_eigenfuns111}Hole subband wavefunctions for growth direction [111]. The four components $\Psi_{1,2,3,4}(r)$ of the wavefunctions Eq.~(\ref{Eq_eigenfunc}) plotted as a function of the coordinate $r$.  The ground (a), the first excited (b), and the second excited (c) states at $k_{z}R=0$. The ground (d), the first excited (e), and the second excited (f) states at $k_{z}R=0.4$. Here, we only consider eigenfunctions with positive total angular momentum $F_{z}>0$.}
\end{figure}

The functional forms of the subband wavefunctions are shown in Fig.~\ref{fig_eigenfuns111}. At $k_{z}R=0$, the wavefunctions always have two non-zero components. Figures~\ref{fig_eigenfuns111}(a), (b) and (c) show the functional forms of the ground, the first excited, and the second excited states at $k_{z}R=0$, respectively. All the four components of the wavefunction are non-zero at a general site $k_{z}R\neq0$. Figures.~\ref{fig_eigenfuns111}(d), (e) and (f) show the functional form of the ground, the first excited, and the second excited states at the site $k_{z}R=0.4$.  

Let us focus on the lowest two eigenstates at the anticrossing site $k_{z}R=0$, i.e., the site with an energy gap. For the growth direction [111], we have the ground state $\Psi_{1,3}(r)\neq0$ and the first excited state $\Psi_{2,4}(r)\neq0$ [see Figs.~\ref{fig_eigenfuns111}(a) and (b)]. While for the growth direction [001], we have the ground state $\Psi_{2,4}(r)\neq0$ and the first excited state $\Psi_{1,3}(r)\neq0$ instead [see Figs.~\ref{fig_subbandwfs001}(a) and (b)].
Hence, at the site of the anticrossing, the eigenstates for growth direction [111] are inverted if we regard the eigenstates for the growth direction [001] as normal. This result induces interesting consequence we will discuss in Sec.~\ref{Sec_VI}.

\section{\label{Sec_V}The effect of the non-axial term}

In this section, we discuss the impact of the non-axial term in the general Lutinger-Kohn Hamiltonian on the hole subband dispersions both for growth directions [001] and [111]. For nanowire growth direction [001], the difference between the general Lutinger-Kohn Hamiltonian and that in the axial approximation reads~\cite{PhysRevB.52.11132,BANGERT1985363}
\begin{equation}
H'_{001}=\frac{\sqrt{3}\hbar^{2}}{4m}(\gamma_{3}-\gamma_{2})\left(\begin{array}{cccc}0&0&k^{2}_{+}&0\\0&0&0&k^{2}_{+}\\k^{2}_{-}&0&0&0\\0&k^{2}_{-}&0&0\end{array}\right).\label{eq_H001}
\end{equation}
For nanowire growth direction [111], the difference between the general Lutinger-Kohn Hamiltonian and that in the axial approximation reads~\cite{PhysRevB.85.235308,PhysRevB.108.165301}
\begin{equation}
H'_{111}=-\frac{(\gamma_{3}-\gamma_{2})\hbar^{2}}{\sqrt{6}m}\left(\begin{array}{cccc}0&k^{2}_{+}&2k_{z}k_{+}&0\\k^{2}_{-}&0&0&2k_{z}k_{+}\\2k_{z}k_{-}&0&0&-k^{2}_{+}\\0&2k_{z}k_{-}&-k^{2}_{-}&0\end{array}\right).\label{eq_H111}
\end{equation} 

We now consider $H'_{001}$ or $H'_{111}$ as a perturbation term~\cite{landau1965quantum}, and discuss its energy correction to the lowest hole subband dispersion shown in Fig.~\ref{fig_subbands001} or \ref{fig_subbands111}. The general lowest subband wavefunction of total angular momentum $F_{z}=1/2$ can be written as~\cite{RL2021}
\begin{equation}
|\Psi_{1/2}\rangle=\left(\begin{array}{c}\Psi_{1}(r)e^{-i\varphi}\\\Psi_{2}(r)\\\Psi_{3}(r)e^{i\varphi}\\\Psi_{4}(r)e^{2i\varphi}\end{array}\right).
\end{equation}
Acting the perturbation Hamiltonian $H'_{001}$ on this state, we have (up to a constant factor)
\begin{equation}
H'_{001}|\Psi_{1/2}\rangle\sim\left(\begin{array}{c}k^{2}_{+}\Psi_{3}(r)e^{i\varphi}\\k^{2}_{+}\Psi_{4}(r)e^{2i\varphi}\\k^{2}_{-}\Psi_{1}(r)e^{-i\varphi}\\k^{2}_{-}\Psi_{2}(r)\end{array}\right).
\end{equation}
By a simple check of the integral of the angular variable $\varphi$ in the perturbation matrix element $\langle\Psi_{F_{z}}|H'_{001}|\Psi_{1/2}\rangle$ by using $\frac{1}{2\pi}\int^{2\pi}_{0}d\varphi{\rm exp}[i(m-m')\varphi]=\delta_{mm'}$, we find the first non-zero perturbation matrix element comes from subband wavefunction of $|F_{z}|=7/2$.

Acting the perturbation Hamiltonian $H'_{111}$ on the state $|\Psi_{1/2}\rangle$, we have (up to a constant factor)
\begin{equation}
H'_{111}|\Psi_{1/2}\rangle\sim\left(\begin{array}{c}k^{2}_{+}\Psi_{2}(r)+2k_{z}k_{+}\Psi_{3}(r)e^{i\varphi}\\k^{2}_{-}\Psi_{1}(r)e^{-i\varphi}+2k_{z}k_{+}\Psi_{4}(r)e^{2i\varphi}\\2k_{z}k_{-}\Psi_{1}(r)e^{-i\varphi}-k^{2}_{+}\Psi_{4}(r)e^{2i\varphi}\\2k_{z}k_{-}\Psi_{2}(r)-k^{2}_{-}\Psi_{3}(r)e^{i\varphi}\end{array}\right).
\end{equation}
By a similar check of the integral of the angular variable $\varphi$ in $\langle\Psi_{F_{z}}|H'_{001}|\Psi_{1/2}\rangle$, we find the first non-zero perturbation matrix element comes from the subband wavefunction of $|F_{z}|=5/2$.

Therefore, if we are interested only in the low-energy suband dispersions, e.g., the two shifted parabolic curves shown in Figs.~\ref{fig_subbands001} and \ref{fig_subbands111}, the perturbation contributions from the non-axial term in the general Luttinger-Kohn Hamiltonian are negligible. Because these two subband dispersions are well separated from those subband dispersions labeled with $|F_{z}|\ge5/2$ or $|F_{z}|\ge7/2$. Note that for other low-symmetry growth directions such as [110] and [112], the difference between the general Luttinger-Kohn Hamiltonian and that in the axial approximation is not as simple as that given by Eqs.~(\ref{eq_H001}) and (\ref{eq_H111})~\cite{PhysRevB.52.11132}. Hence, for these growth directions, the perturbation corrections from the non-axial term to the low-energy hole subband dispersions may no longer be negligible.

\section{\label{Sec_VI}The gap closes at a special growth direction}

\begin{figure}
\includegraphics{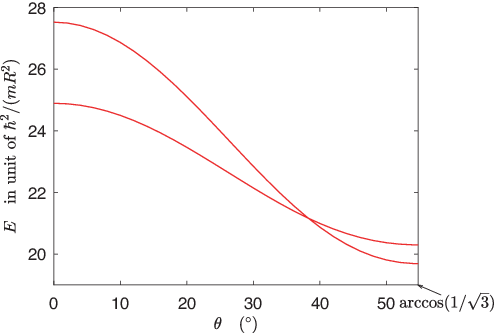}
\caption{\label{fig_levelcrossing}The lowest two subband energies at $k_{z}R=0$ as a function of the nanowire growth direction $\theta$. Here $\phi$ is fixed at $\phi=45^{\circ}$. $\theta=0^{\circ}$ corresponds to the growth direction [001] and $\theta=\arccos(1/\sqrt{3})$ corresponds to the growth direction [111]. The energy gap closes at $\theta\approx38.258^{\circ}$.}
\end{figure}

From the analysis given in Sec.~\ref{Sec_V}, we believe that the axial approximation indeed reasonably gives rise to both the low-energy subband dispersions and the corresponding subband wavefunctions both for growth directions [001] and [111]. Here, we discuss the possibility of a gap closing at a special growth direction. 
 
The lowest two eigenstates at $k_{z}R=0$ for growth direction [111] are inverted in comparison with the normal eigenstates for growth direction [001]. Quantum state invertion has been extensively studied in the field of topological insulator~\cite{doi:10.1126/science.1133734,RevModPhys.83.1057}, and state invertion usually indicates the existence of a gap closing site. Here, we search this gap closing site between nanowire growth directions [001] and [111]. 

When one angle of the growth direction is fixed at $\phi=45^{\circ}$, and the other angle $\theta$ is varied from $0^{\circ}$ to $\arccos(1/\sqrt{3})$, i.e., from the growth direction [001]  to the growth direction [111], the lowest two subband energies at the site $k_{z}R=0$ as a function of $\theta$ are shown in Fig.~\ref{fig_levelcrossing}. We find that there indeed exists a gap closing site at $\theta\approx38.258^{\circ}$. Although we obtain this gap closing site by using the Luttinger-Kohn Hamiltonian in the axial approximation, we believe a gap closing site still exists even if the general Luttinger-Kohn Hamiltonian is used. 

\begin{figure}
\includegraphics{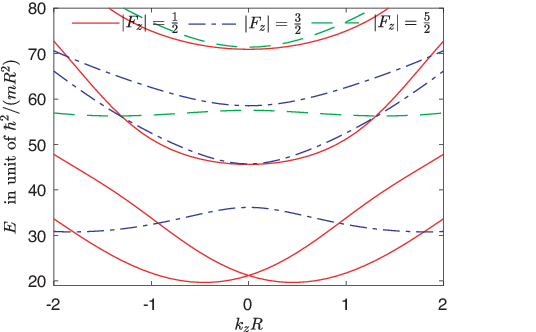}
\caption{\label{fig_subbandcritical}The hole subband dispersions for the nanowire growth direction $\theta=38.258^{\circ}$ and $\phi=45^{\circ}$. The band minima are located at $k_{z}R=\pm0.5$. Each dispersion line is two-fold degenerate, i.e., spin degenerate. There is no energy gap at $k_{z}R=0$ for the lowest two subband dispersions.}
\end{figure}

The hole subband dispersions at the critical growth direction $\theta=38.258^{\circ}$ and $\phi=45^{\circ}$ are shown in Fig.~\ref{fig_subbandcritical}. The lowest two subband dispersions now are two shifted parabolic curves with no energy gap at $k_{z}R=0$, i.e., the lowest two parabolic curves cross with each other at $k_{z}R=0$. The subband minima are located at $k_{z}R\approx\pm0.5$.

\section{Discussion and summary}
{\color{red}}
The effective Hamiltonian describing the lowest two hole subband dispersions, i.e., the two shifted parabolic  curves with an anticrossing at $k_{z}R=0$, can be written as~\cite{PhysRevB.84.195314,RL2023b}
\begin{equation}
H^{\rm ef}\approx\frac{\hbar^{2}k^{2}_{z}}{2m^{*}_{h}}+\alpha\,k_{z}\tau^{x}s^{x}+\frac{\Delta}{2}\tau^{z},\label{eq_Hamiltoniankz2}
\end{equation}
where $m^{*}_{h}$ is the effective hole mass, $\alpha$ is the `spin'-orbit coupling strength, $\Delta$ is the energy gap at the anticrossing site $k_{z}R=0$, $\boldsymbol \tau$ is the `spin' (pseudo spin) operator defined in the Hilbert subspace spanned by the lowest two orbital states at $k_{z}R=0$, and ${\bf s}$ is the real hole spin operator. The value of $\Delta$ is positive for growth direction [001] (Fig.~\ref{fig_subbands001}), is negative for growth direction [111] (Fig.~\ref{fig_subbands111}), and is zero for the critical growth direction $\theta\approx38.258^{\circ}$ and $\phi=45^{\circ}$ (Fig.~\ref{fig_subbandcritical}). The low-energy subband dispersions of the hole gas in the nanowire are thus apparently distinct from that of the electron gas, where the subband minima of the electron gas are usually located at the center of the $k_{z}$ space, i.e., $k_{z}R=0$~\cite{gulyamov2020electron,doi:10.1063/1.4929412,doi:10.1063/1.4972987}.

In summary, we have exactly solved the effective mass model of the hole gas in a cylindrical Ge nanowire, where the Luttinger-Kohn Hamiltonian in the axial approximation is used. More general transcendental equations beyond the spherical approximation are analytical derived, and the solutions of these equations give rise to the hole subband dispersions in the nanowire. The axial approximation is almost accurate for high-symmetry nanowire growth directions [001] and [111]. The lowest two subband dispersions for these two growth directions can be regarded as two shifted parabolic curves with an energy gap the $k_{z}R=0$. However, the eigenstates for growth direction [111] at the gap opening site are inverted comparing with the normal eigenstates for growth direction [001]. Quantum state invertion at $k_{z}R=0$ prompts us to search a gap closing growth direction between [001] and [111]. Calculations based on the axial approximation show this critical growth direction is located at $\theta\approx38.258^{\circ}$ and $\phi=45^{\circ}$.

\appendix
\section{\label{app_a}The bulk spectrum and the bulk wavefunctions}
The total Hilbert space can be divided into a series of subspaces by considering the conservation of the total angular momentum $F_{z}$. The Hilbert subspace specified with a general total angular momentum $F_{z}=m+1/2$ is spanned by $J_{m-1}(\mu\,r)e^{i(m-1)\varphi}|3/2\rangle$, $J_{m}(\mu\,r)e^{im\varphi}|1/2\rangle$, $J_{m+1}(\mu\,r)e^{i(m+1)\varphi}|-1/2\rangle$, and $J_{m+2}(\mu\,r)e^{i(m+2)\varphi}|-3/2\rangle$. The bulk Hamiltonian $H^{\rm axial}_{\rm LK}$ can be written as a $4\times4$ matrix in this Hilbert subspace [in unit of $\hbar^{2}/(2m)$]
\begin{equation}
H^{\rm axial}_{\rm LK}=\left(\begin{array}{cccc}H_{11}&2\sqrt{3}i\tilde{\gamma}_{2}k_{z}\mu&\sqrt{3}\tilde{\gamma}_{3}\mu^{2}&0\\
 h.c.&H_{22}&0&\sqrt{3}\tilde{\gamma}_{3}\mu^{2}\\
h.c.&0&H_{33}&-2\sqrt{3}i\tilde{\gamma}_{2}k_{z}\mu\\
0&h.c.&h.c.&H_{44}\end{array}\right),\label{eq_app1}
\end{equation}
%\begin{widetext}
%\begin{equation}
%H^{\rm axial}_{\rm LK}=\frac{\hbar^{2}}{2m}\left(\begin{array}{cccc}H_{11}&2\sqrt{3}i\tilde{\gamma}_{2}k_{z}\mu&\sqrt{3}\tilde{\gamma}_{3}\mu^{2}&0\\
%-2\sqrt{3}i\tilde{\gamma}_{2}k_{z}\mu&H_{22}&0&\sqrt{3}\tilde{\gamma}_{3}\mu^{2}\\
%\sqrt{3}\tilde{\gamma}_{3}\mu^{2}&0&H_{33}&-2\sqrt{3}i\tilde{\gamma}_{2}k_{z}\mu\\0&\sqrt{3}\tilde{\gamma}_{3}\mu^{2}&2\sqrt{3}i\tilde{\gamma}_{2}k_{z}\mu&H_{44}\end{array}\right),\label{eq_app1}
%\end{equation}
%\end{widetext}
where $H_{11}=H_{44}=(\gamma_{1}+\tilde{\gamma}_{1})\mu^{2}+(\gamma_{1}-2\tilde{\gamma}_{1})k^{2}_{z}$, $H_{22}=H_{33}=(\gamma_{1}-\tilde{\gamma}_{1})\mu^{2}+(\gamma_{1}+2\tilde{\gamma}_{1})k^{2}_{z}$, and $h.c.$ denotes the Hermitian conjugation. Diagonalizing the matrix (\ref{eq_app1}), we obtain two branchs of bulk spectrum
\begin{equation}
E=\frac{\hbar^{2}}{2m}[\gamma_{1}(\mu^{2}+k^{2}_{z})\pm\,X_{\mu}],
\end{equation} 
where $X_{\mu}$ is given in Eq.~(\ref{eq_Xmu}). There are two bulk wavefunctions corresponding to the `+' branch spectrum
\begin{eqnarray}
\left(\begin{array}{c}\frac{2i\tilde{\gamma}_{2}k_{z}}{\tilde{\gamma}_{3}\mu}J_{m-1}(\mu\,r)e^{i(m-1)\varphi}\\\frac{-\tilde{\gamma}_{1}(\mu^{2}-2k^{2}_{z})+X_{\mu}}{\sqrt{3}\tilde{\gamma}_{3}\mu^{2}}J_{m}(\mu\,r)e^{im\varphi}\\
0\\
J_{m+2}(\mu\,r)e^{i(m+2)\varphi}\end{array}\right),\label{eq_bulkstateplus1}
\end{eqnarray}
and
\begin{equation}
\left(\begin{array}{c}
\frac{\tilde{\gamma}_{1}(\mu^{2}-2k^{2}_{z})+X_{\mu}}{\sqrt{3}\tilde{\gamma}_{3}\mu^{2}}J_{m-1}(\mu\,r)e^{i(m-1)\varphi}\\
-\frac{2i\tilde{\gamma}_{2}k_{z}}{\tilde{\gamma}_{3}\mu}J_{m}(\mu\,r)e^{im\varphi}\\
J_{m+1}(\mu\,r)e^{i(m+1)\varphi}\\
0
\end{array}\right).\label{eq_bulkstateplus2}
\end{equation}
There are also two bulk wavefunctions corresponding to the `-' branch spectrum
\begin{eqnarray}
\left(\begin{array}{c}\frac{2i\tilde{\gamma}_{2}k_{z}}{\tilde{\gamma}_{3}\mu}J_{m-1}(\mu\,r)e^{i(m-1)\varphi}\\\frac{-\tilde{\gamma}_{1}(\mu^{2}-2k^{2}_{z})-X_{\mu}}{\sqrt{3}\tilde{\gamma}_{3}\mu^{2}}J_{m}(\mu\,r)e^{im\varphi}\\
0\\
J_{m+2}(\mu\,r)e^{i(m+2)\varphi}\end{array}\right),\label{eq_bulkstateminus1}
\end{eqnarray}
and
\begin{equation}
\left(\begin{array}{c}
\frac{\tilde{\gamma}_{1}(\mu^{2}-2k^{2}_{z})-X_{\mu}}{\sqrt{3}\tilde{\gamma}_{3}\mu^{2}}J_{m-1}(\mu\,r)e^{i(m-1)\varphi}\\
-\frac{2i\tilde{\gamma}_{2}k_{z}}{\tilde{\gamma}_{3}\mu}J_{m}(\mu\,r)e^{im\varphi}\\
J_{m+1}(\mu\,r)e^{i(m+1)\varphi}\\
0
\end{array}\right).\label{eq_bulkstateminus2}
\end{equation}

\section{\label{app_b}The equation array for the coefficients $c_{1,2,3,4}$}
The hard-wall boundary condition $\Psi^{\rm I/II}(R,\varphi,z)=0$ can be written as
\begin{widetext}
\begin{eqnarray}
\frac{2i\tilde{\gamma}_{2}k_{z}}{\tilde{\gamma}_{3}\mu_{1}}J_{m-1}(\mu_{1}R)c_{1}+\frac{\tilde{\gamma}_{1}(\mu^{2}_{1}-2k^{2}_{z})\pm\,X_{\mu_{1}}}{\sqrt{3}\tilde{\gamma}_{3}\mu^{2}_{1}}J_{m-1}(\mu_{1}R)c_{2}+\frac{2i\tilde{\gamma}_{2}k_{z}}{\tilde{\gamma}_{3}\mu_{2}}J_{m-1}(\mu_{2}R)c_{3}+\frac{\tilde{\gamma}_{1}(\mu^{2}_{2}-2k^{2}_{z})-X_{\mu_{2}}}{\sqrt{3}\tilde{\gamma}_{3}\mu^{2}_{2}}J_{m-1}(\mu_{2}R)c_{4}&=&0,\nonumber\\
\frac{-\tilde{\gamma}_{1}(\mu^{2}_{1}-2k^{2}_{z})\pm\,X_{\mu_{1}}}{\sqrt{3}\tilde{\gamma}_{3}\mu^{2}_{1}}J_{m}(\mu_{1}R)c_{1}-\frac{2i\tilde{\gamma}_{2}k_{z}}{\tilde{\gamma}_{3}\mu_{1}}J_{m}(\mu_{1}R)c_{2}-\frac{\tilde{\gamma}_{1}(\mu^{2}_{2}-2k^{2}_{z})+X_{\mu_{2}}}{\sqrt{3}\tilde{\gamma}_{3}\mu^{2}_{2}}J_{m}(\mu_{2}R)c_{3}-\frac{2i\tilde{\gamma}_{2}k_{z}}{\tilde{\gamma}_{3}\mu_{2}}J_{m}(\mu_{2}R)c_{4}&=&0,\nonumber\\
J_{m+1}(\mu_{1}R)c_{2}+J_{m+1}(\mu_{2}R)c_{4}&=&0,\nonumber\\
J_{m+2}(\mu_{1}R)c_{1}+J_{m+2}(\mu_{2}R)c_{3}&=&0.\nonumber\\
\end{eqnarray}
\end{widetext}

\section{\label{app_c}Lowest subband dispersions both of $|F_{z}|=3/2$ and $|F_{z}|=5/2$ for growth direction [001]}
\begin{figure}
\includegraphics{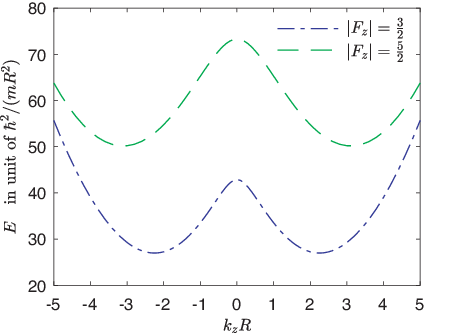}
\caption{\label{fig_subbands001_long}The lowest subband dispersions both of $|F_{z}|=3/2$ and $|F_{z}|=5/2$ for growth direction [001].}
\end{figure}
In Fig.~\ref{fig_subbands001}, we only show subband dispersions for wave vectors up to $|k_{z}R|=2$, where the minima of the lowest subband dispersions both of $|F_{z}|=3/2$ and $|F_{z}|=5/2$ are not covered. Here we calculate  these two subband dispersions for wave vectors up to $|k_{z}R|=5$ (see Fig.~\ref{fig_subbands001_long}), where the subband minima can be clearly seen now. The minima of the subband dispersion $|F_{z}|=3/2$ have energy [about $26.99\hbar^{2}/(mR^{2})$] higher than that of the subband dispersion $|F_{z}|=1/2$ [about $22.76\hbar^{2}/(mR^{2})$], such that the minima of the lowest suband dispersion of $|F_{z}|=1/2$ are actually the band minima.

\bibliography{Ref_Hole_spin}
\end{document}